\documentclass[%
 reprint,
 superscriptaddress,
 amsmath,amssymb,
 aps,
 prl,
]{revtex4-2}
\usepackage{graphicx}
\usepackage{dcolumn}
\usepackage{bm}

\usepackage{mathtools}
\usepackage{float}
\usepackage[usenames,dvipsnames]{color}
\usepackage[colorlinks=true, citecolor=blue, linkcolor=blue, urlcolor=blue]{hyperref}
\usepackage{xr-hyper}
\usepackage{bm}
\usepackage{cleveref} 
\usepackage[printonlyused]{acronym}
\usepackage{layouts}

\makeatletter
\newcommand*{\addFileDependency}[1]{
  \typeout{(#1)}
  \@addtofilelist{#1}
  \IfFileExists{#1}{}{\typeout{No file #1.}}
}
\makeatother

\newcommand*{\myexternaldocument}[1]{%
    \externaldocument{#1}%
    \addFileDependency{#1.tex}%
    \addFileDependency{#1.aux}%
}

\myexternaldocument{Supplementary}

\newcommand{\ada}{a^{\dagger}a}

\newcommand{\aplusa}{a^{\dagger} {+} a}

\begin{document}

\title{Cavity Light-Matter Entanglement through Quantum Fluctuations}
\newcommand{\affiliationRWTH}{
Institut f\"ur Theorie der Statistischen Physik, RWTH Aachen University and JARA-Fundamentals of Future Information Technology, 52056 Aachen, Germany
}
\newcommand{\affiliationMPSD}{
Max Planck Institute for the Structure and Dynamics of Matter,
Center for Free-Electron Laser Science (CFEL),
Luruper Chaussee 149, 22761 Hamburg, Germany
}

\newcommand{\affiliationBristol}{
H H Wills Physics Laboratory, University of Bristol, Bristol BS8 1TL, United
Kingdom 
}

\newcommand{\affiliationUNIGE}{
Dipartimento di Fisica, Università di Genova, 16146, Genova, Italy
}
\newcommand{\affiliationCNR}{
SPIN-CNR, 16146, Genova, Italy
}

\author{Giacomo Passetti}
\thanks{These authors contributed equally}
\affiliation{\affiliationRWTH}

\author{Christian J. Eckhardt}
\thanks{These authors contributed equally}
\affiliation{\affiliationMPSD}
\affiliation{\affiliationRWTH}



\author{Michael~A. Sentef}
\affiliation{\affiliationBristol}\affiliation{\affiliationMPSD}

\author{Dante M.~Kennes}
\affiliation{\affiliationRWTH}
\affiliation{\affiliationMPSD}

\date{\today}
\begin{abstract}
\noindent The hybridization between light and matter forms the basis to achieve cavity control over quantum materials.
In this work we investigate a cavity coupled to an XXZ quantum chain of interacting spinless fermions by numerically exact solutions and perturbative analytical expansions. We find two important effects: (i) Specific quantum fluctuations of the matter system play a pivotal role in achieving entanglement between light and matter; and (ii) in turn, light-matter entanglement is the key ingredient to modify electronic properties by the cavity. We hypothesize that quantum fluctuations of those matter operators to which the cavity modes couple are a general prerequisite for light-matter entanglement in the groundstate. Implications of our findings for light-matter-entangled phases, cavity-modified phase transitions in correlated systems, and measurement of light-matter entanglement through Kubo response functions are discussed.

\end{abstract}
\maketitle


\paragraph*{Introduction.--}\label{INTRO}
Controlling material properties with light is a tantalizing avenue in condensed matter physics \cite{de_la_torre_colloquium_2021}.
Notable recent achievements based on the ultrafast interaction of laser pulses with quantum materials include the light-induced anomalous Hall effect in graphene \cite{mciver_light-induced_2020} and nonequilibrium superconducting-like states \cite{mitrano_possible_2016,buzzi_photomolecular_2020}.
Going beyond classical electromagnetic fields, materials properties can be influenced by hybridizing them strongly with the quantized electromagnetic field in optical resonators.
An early example is the Purcell effect, where a change of the structure of the electromagnetic vacuum leads to a modified rate of spontaneous emission, or even Rabi oscillations \cite{purcell_spontaneous_1946}.
More recently, the control of vacuum fluctuations and light-matter hybridization are utilized to design properties of extended solids, giving rise to a research field that has been coined ``cavity quantum materials'' \cite{ruggenthaler_quantum-electrodynamical_2018, frisk_kockum_ultrastrong_2019, schlawin_cavity_2022, hubener_engineering_2021}.
Here coupling between quantum materials and cavity modes in a variety of cavity types and geometries can lead to the formation of hybrid polaritonic light-matter states with sometimes drastically modified many-body properties controlled by designing the electromagnetic environment \cite{appugliese_breakdown_2022, ciuti_cavity-mediated_2021, thomas_ground-state_2016, schafer_shining_2022, orgiu_conductivity_2015, hagenmuller_cavity-enhanced_2017, bhatt_enhanced_2021, paravicini-bagliani_magneto-transport_2019, thomas_exploring_2019, schlawin_cavity-mediated_2019, mazza_superradiant_2019, gao_photoinduced_2020, chakraborty_long-range_2021, sentef_cavity_2018, andolina_can_2022}. 

The entanglement entropy between the light and matter parts of a system can be used as an indicator for the degree of their hybridization.
Therefore the creation of light-matter entanglement is one of the key resources in the field.
In the case of plain-vanilla polaritons, namely hybridized bosonic modes, \cite{basov_polariton_2021} the generation of entanglement can be understood through models of coupled oscillators \cite{lambert_entanglement_2004}. However in the growing field of (correlated) band electrons coupled to a cavity \cite{jarc_cavity_2022, rokaj_free_2022, guerci_superradiant_2020,
Andolina2020, Andolina_nogo,
chiriaco_critical_2022, eckhardt_quantum_2021, chiriaco_critical_2022, gao_photoinduced_2020, schlawin_cavity-mediated_2019, rao_non-fermi-liquid_2022, chakraborty_long-range_2021, chakraborty_controlling_2022, latini_ferroelectric_2021, li_effective_2022} a clear picture of necessary and sufficient conditions for light-matter entanglement is still missing.

In this Letter we remedy this long-standing problem by showing that the quantum fluctuations of the current operator, which is generically the matter operator to which photonic gauge fields couple, play a pivotal role for achieving entanglement between correlated electrons and a cavity.
By studying a paradigmatic and experimentally relevant model, we conclude that such fluctuations in fact constitute a necessary condition for non-zero light-matter entanglement.
Additionally we find that a mean-field decoupling of light and matter, thus neglecting entanglement, yields qualitatively wrong results for photon related observables and does not reproduce the dominant corrections to fermionic correlation functions.


\paragraph*{Model.--}
\begin{figure*}
    \centering
    \includegraphics{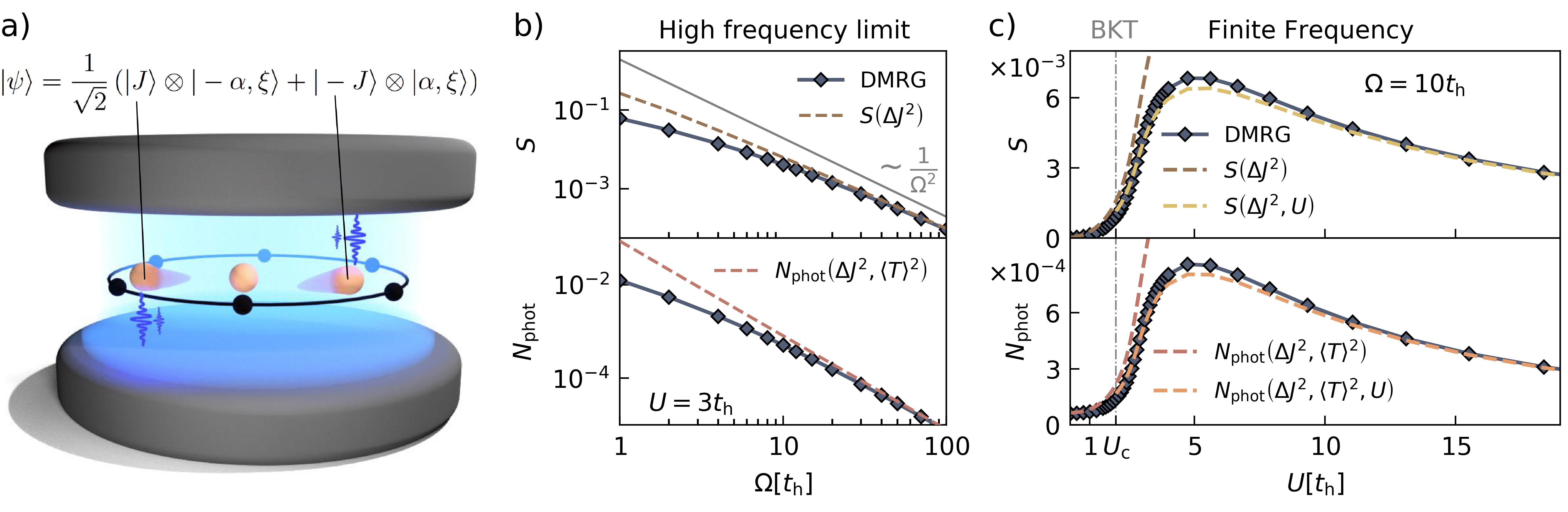}
    \caption{\textbf{a)} Illustration of the mechanism generating light-matter entanglement in the ground state.
    While the expectation value of the current always vanishes $\langle J \rangle = 0$, the system is in a super-position of Bloch states with a well defined eigenvalue of the current operator $J$ ($|\pm J\rangle$, only two are noted here for illustration) that each come with a certain squeezed coherent state $|\mp \alpha, \xi \rangle$ of the cavity mode Eq.~(\ref{eq:approxWavefunction}) where the product $J \alpha$ has a fixed sign such that the overall inversion symmetry of the system is preserved. 
    \textbf{b)} Entanglement entropy $S$ (top) and photon number $N_{\rm phot}$ (bottom) as computed with DMRG (blue markers) as a function of the cavity frequency $\Omega$; and as given by Eq.~(\ref{eq:S_j}) $S(\Delta J^2)$ (top, brown dashed line) and Eq.~(\ref{eq:nPhotHighFreq}) $N_{\rm phot}(\Delta J^2, \langle T \rangle^2)$ (bottom, red dashed line) in the high frequency limit.
    The gray line marks a $\frac{1}{\Omega^2}$ decay and we set $U=3t_{\rm h}$ for the fermion-fermion interaction.
    \textbf{c)} Entanglement entropy $S$ (top) and photon number $N_{\rm phot}$ (bottom) obtained with DMRG (blue marks) as function of $U$.
    Expressions obtained in the high-frequency limit Eq.~(\ref{eq:S_j}) $S(\Delta J^2)$ (top, brown dashed line) and Eq.~(\ref{eq:nPhotHighFreq}) $N_{\rm phot}(\Delta J^2, \langle T \rangle^2)$ (bottom, red dashed line) as well as those accounting for $U \sim \Omega$ Eq.~(\ref{eq:S_jU}) $S(\Delta J^2, U)$ (top, yellow dashed line) and Eq.~(\ref{eq:nPhotHighFreqU}) $N_{\rm phot}(\Delta J^2, \langle T \rangle^2, U)$ (bottom, orange dashed line) are overlayed.
    The frequency is set to $\Omega = 10t_{\rm h}$ in \textbf{c)}, the chain length to $L=110$ and the light-matter interaction $g = 0.5$ throughout.
    }
    \label{fig:1}
\end{figure*}
We consider a model of interacting spinless fermions
coupled to the first transmittance resonance of a cavity fixing the wave vector perpendicular to the chain. 
In the direction of the chain, that we take as the $x$-direction, we only consider the spatially constant mode, i.e., wave vector $q_x = 0$.
This amounts to the dipole approximation, which is justified for optical cavities since $c \gg v_{\rm F}$, where $v_{\rm F}$ is the Fermi velocity.
We employ a quantized version of the Peierls substitution to couple light and matter in Coulomb gauge. This guarantees a gauge-invariant coupling for a low-energy theory \cite{li_electromagnetic_2020, dmytruk_gauge_2021},
\begin{equation}
\begin{aligned}
    H &= -t_{h}\sum_{j=1}^{L}\left(e^{i\frac{g}{\sqrt{L}}(\aplusa)}c^{\dag}_{j}c_{j+1}+h.c.\right) +\\
    &U\sum_{j=1}^{L}\left(n_{j}-\frac{1}{2}\right)\left(n_{j+1}-\frac{1}{2}\right) + \Omega \ada .
    \label{eq:HFull}
\end{aligned}
\end{equation}

Here, $c_j^{(\dag)}$ annihilates (creates) a fermion at site $j$, and $n_j = c_j^{\dag} c_j$ is the corresponding fermionic occupation.
$a^{(\dag)}$ annihilates (creates) a cavity photon, $\Omega$ denotes the bare cavity frequency, and $g$ parametrizes the light-matter coupling.
$t_h$ denotes the nearest-neighbor hopping amplitude, $U$ the nearest neighbor repulsion, and $L$ the total number of sites.

We use periodic boundary conditions and a half-filled band.
We set $e = \hbar =\rm{k}_{\rm B} =   c = 1 $ and the lattice constant $a_{\rm lat}=1$.
For $g = 0$ this model is equivalent to the XXZ spin Hamiltonian for which the exact groundstate can be obtained with the Bethe Ansatz \cite{giamarchi_quantum_2003}.
At $U = 2t_{\rm h}$ the fermions undergo a quantum phase transition of the Berezinksii-Kosterlitz-Thouless (BKT) type from a Luttinger liquid (LL) phase to a charge-density wave (CDW) \cite{campos_venuti_quantum_2007, sun_fidelity_2015, chen_intrinsic_2008, li_entanglement_2016, du_visualizing_2021}.
For $g>0$ the fermionically noninteracting limit ($U = 0$) can be solved analytically and has a groundstate given by the product state of the unchanged Fermi sea for the fermions and a squeezed state for the cavity photon mode \cite{eckhardt_quantum_2021}.

\paragraph*{Analytical results.--}
First, we expand the Hamiltonian Eq.~(\ref{eq:HFull}) to quadratic order in $g$.
Employing a combined displacement and squeezing transformation of the Hamiltonian (see Supplemental Material \cite{Supplementary}), we find a formal expression for the groundstate wave function of the system that is controlled in a high-frequency expansion,
\begin{equation}
    \begin{aligned}
     |\Psi \rangle = \sum_u c_u | j_u, t_u \rangle \otimes |\alpha(j_u), \xi(t_u) \rangle,
    \end{aligned}
    \label{eq:approxWavefunction}
\end{equation}
where $c_u$ are complex coefficients fulfilling $\sum_u |c_u|^2 = 1$, $|j_u, t_u\rangle$ are Bloch states composed of eigenstates of the current operator $J = \sum_k 2 t_{\rm h} \sin(k) c_k^{\dag} c_k$ and kinetic energy operator $T = -\sum_k 2 t_{\rm h} \cos(k) c_k^{\dag} c_k$, obeying $J|j_u\rangle = j_u |j_u\rangle$ and $T|t_u\rangle = t_u |t_u\rangle$, respectively. $|\alpha(j_u), \xi(t_u) \rangle$ are squeezed coherent states with displacement parameter $\alpha(j_u) = -\frac{g}{\sqrt{L}\Omega} j_u$ and squeezing parameter $\xi(t_u) = \frac{1}{2} \ln \left(1 - 2\frac{g^2}{\Omega} \frac{t_u}{L} \right)$. 
The expression Eq.~(\ref{eq:approxWavefunction}) allows us to formally relate the photon number to expectation values of fermionic operators in the high-frequency limit
\begin{equation}
N_{\rm phot}(\Delta J^2, \langle T \rangle^2) = \langle \Psi | a^{\dag} a | \Psi \rangle = \frac{g^2}{\Omega^2} \frac{\Delta J^2}{L} + \frac{g^4}{\Omega^2} \frac{\langle T \rangle^2}{L^2}.
    \label{eq:nPhotHighFreq}
\end{equation}
To compute the entanglement entropy between light and matter we formally write the reduced density matrix of the photons $\rho_{\rm ph}$ by performing a partial trace of the fermionic degrees of freedom $\text{Tr}_{\rm{f}}$ on the full density matrix $\rho$, yielding
\begin{equation}
    \rho_{\rm ph} = \text{Tr}_{\rm{f}} \, \rho = \sum_u |c_u|^2 | \alpha(j_u), \xi(t_u) \rangle \langle \alpha(j_u), \xi(t_u) |.
    \label{eq:rhoPhot}
\end{equation}
The corresponding entropy $S = - \rm{Tr_{\rm ph}} \, \rho_{\rm ph} \ln(\rho_{\rm ph})$ quantifies the entanglement between light and matter.
In the Supplemental Material \cite{Supplementary} we show $S[\rho_{\rm ph}] \stackrel{L \to \infty}{\longrightarrow} S[\rho'_{\rm ph}] $ where
\begin{equation}
    \rho'_{\rm ph} = \sum_u |c_u|^2 |\alpha(j_u) \rangle \langle \alpha(j_u)|,
    \label{eq:rhoD}
\end{equation}
i.e., the squeezing does not contribute to the entanglement in the thermodynamic limit.
In the case of vanishing current fluctuations $\Delta J^2 = 0$ only a single coefficient $c_u$ in Eq.~(\ref{eq:rhoD}) is different from zero leading to the reduced photonic density matrix of a pure state with vanishing entropy.
We thus conclude the key results
\begin{equation}
    \Delta J^2 = 0 \; \Rightarrow \; S = 0.
    \label{eq:Implication1}
\end{equation}
Hence current fluctuations are a necessary condition for light-matter entanglement.
Computing the entanglement entropy between light and matter in the high frequency limit to leading order in $\frac{1}{\Omega}$ yields
\begin{equation}
\begin{aligned}
    S(\Delta J^2) =& -\frac{1}{1 + \chi} \ln \left(\frac{1}{1 +\chi} \right) -\frac{\chi}{1 + \chi} \ln \left(\frac{\chi}{1 + \chi} \right)\\[8pt]
    \chi =& \frac{g^2}{\Omega^2} \frac{\Delta J^2 }{L}.
\end{aligned}
    \label{eq:S_j}
\end{equation}
We note that $\chi = \frac{g^2}{\Omega^2} \frac{\Delta J^2 }{L} \stackrel{L \to \infty}{\longrightarrow}  \text{const} $ has a non-zero value in the thermodynamic limit.

\begin{figure*}[t]
  \includegraphics{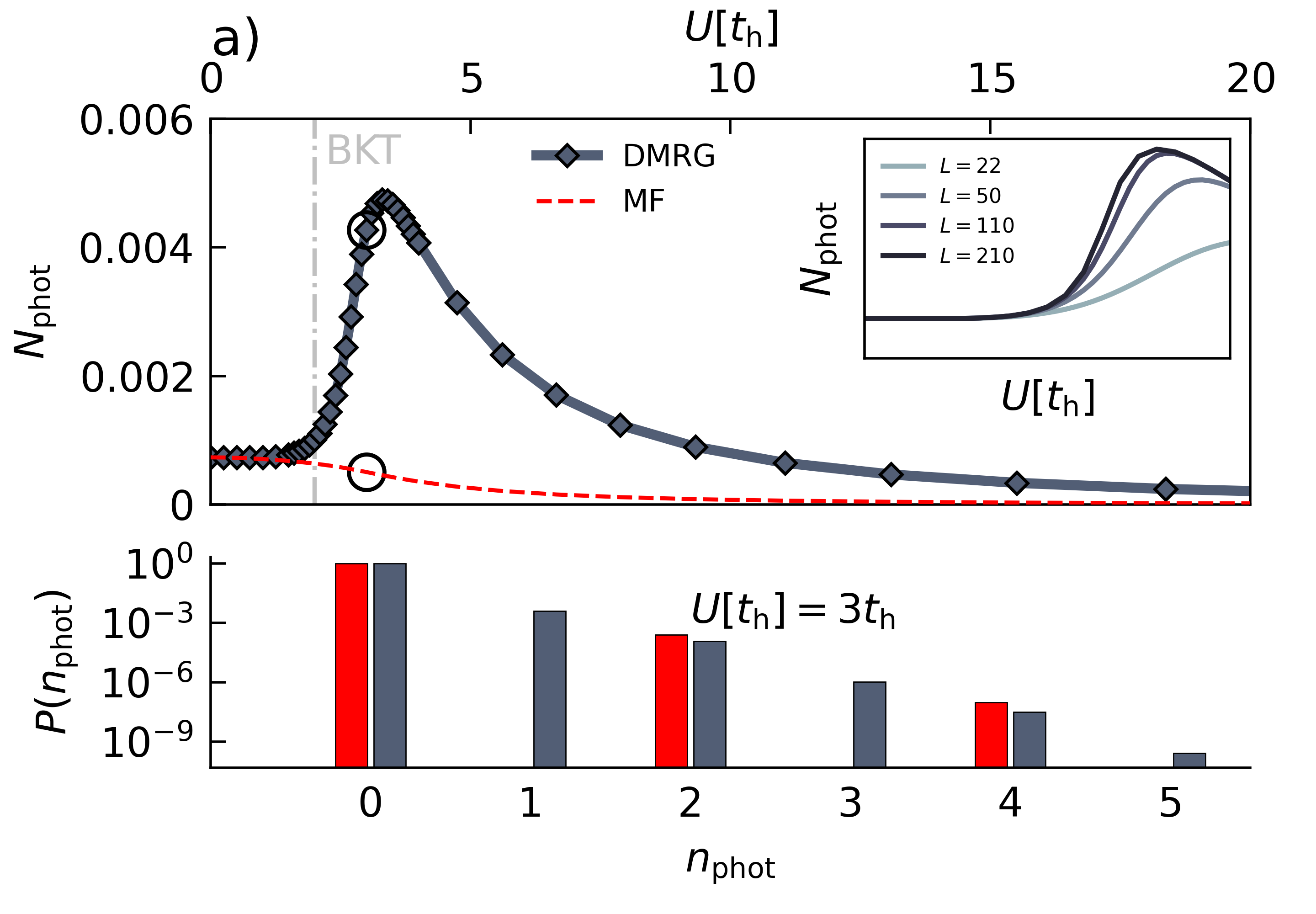}%
  \includegraphics{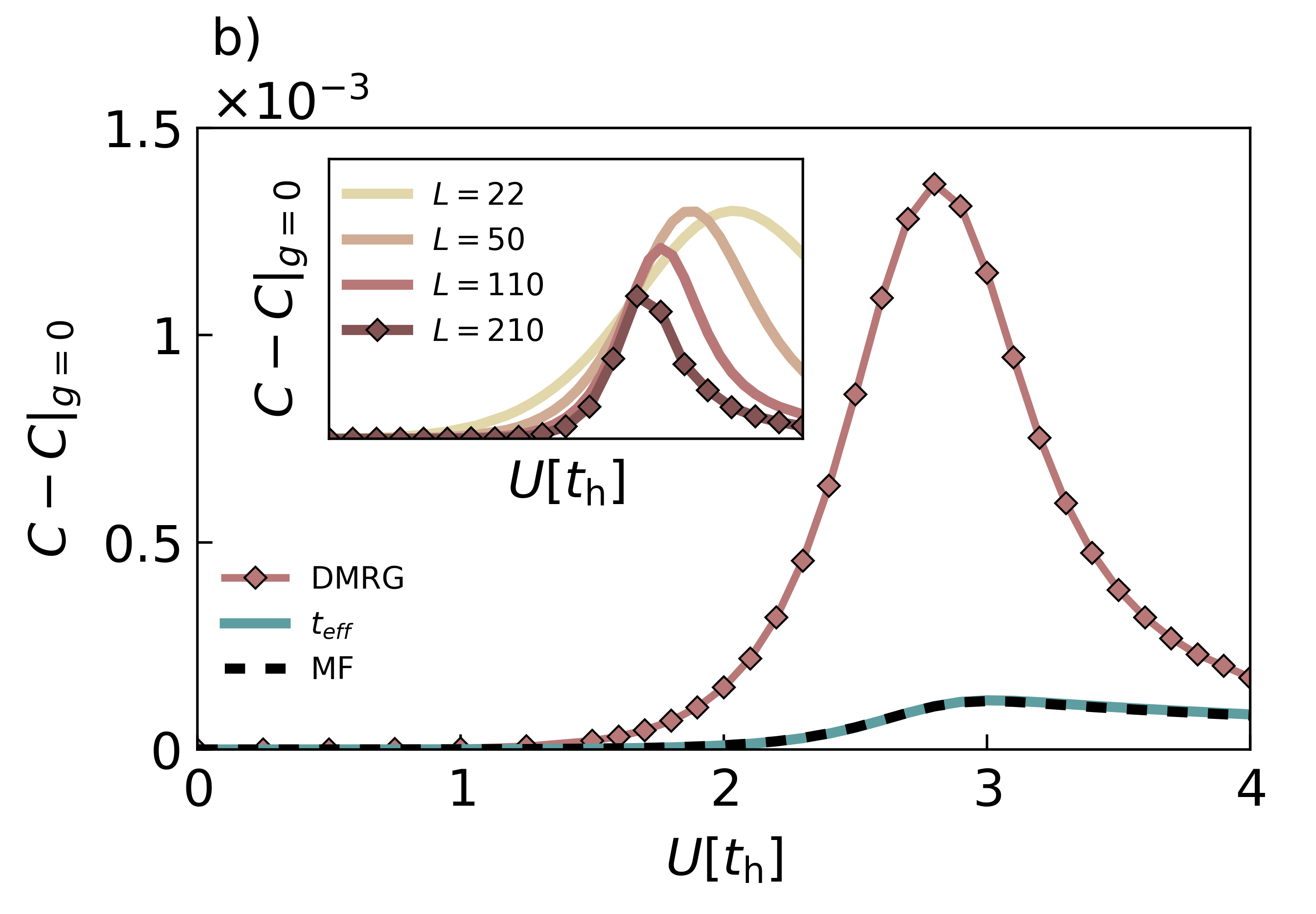}
  \caption{\textbf{a)} Photon number $N_{\rm phot}$ as obtained with DMRG (dark blue points) and with a MF decoupling of light and matter (red dashed line) as a function of interaction strength $U$. The  inset shows a finite size scaling of the photon number in the range $U=0$ to $U=4t_{\rm h}$ as obtained in DMRG.
  The lower plot shows the probability to find $n_{\rm phot}$ photons in the ground state at $U=3t_{\rm h}$ (marked by circles in the upper part) as obtained with DMRG and the MF approach.
  \textbf{b)} Difference of long range correlations $C$ (Eq.~(\ref{eq:def_C})) in the uncoupled system $g=0$ and the coupled one $g = 0.3$ as obtained in DMRG (red points) and the mean field (MF) approach (black dashed line, also see Appendix \ref{sec:mean field decoupling}).
We overlay the expected change of the correlator due to a change of the fermionic hopping $t_{\rm eff}$ (Eq.~(\ref{eq:effectiveHopping})) in blue that precisely matches the result from the MF approach but is at stark odds with the exact results.
The inset shows a finite size scaling for the same quantity as obtained in DMRG.
The chain-length is set to $L=110$ and the frequency to $\Omega = 1t_{\rm h}$.}
\label{fig:2}
\end{figure*}

\paragraph*{Numerical solution.--}
We solve the model Eq.~(\ref{eq:HFull}) including all orders of the gauge-invariant Peierls coupling \cite{eckhardt_quantum_2021, dmytruk_gauge_2021} with the density-matrix renormalization group (DMRG), using a photon number cutoff in the photonic part of the Hilbert space.
For the implementation we use functions from the TeNPy library \cite{hauschild_efficient_2018} and show further numerical details in the Supplemental Material \cite{Supplementary}.

The entanglement entropy $S$ and photon number $N_{\rm phot}$ at $U = 3t_{\rm h}$ and $g = 0.5$ are shown as a function of the bare cavity frequency $\Omega$ in Fig.~\ref{fig:1}b.
For comparison we also show the asymptotic high-frequency result obtained from Eqs.~(\ref{eq:S_j}) and (\ref{eq:nPhotHighFreq}).
We note that the population of the cavity stems solely from modified vacuum fluctuations and hence is small.

We now fix the frequency to an intermediate value of $\Omega = 10t_{\rm h}$.
The light-matter coupling is set to $g=0.5$.
The entanglement entropy as a function of the fermion-fermion interaction $U$ is shown in Fig.~\ref{fig:1}c.
For $U \to 0$ the entanglement between light and matter vanishes together with the current fluctuations.
We compare the result obtained with DMRG to Eq.~(\ref{eq:S_j}) and find  agreement for values of the interaction $U$ up until the point where the interaction becomes comparable to the frequency at around $U = 3t_{\rm h}$.
Beyond that point current fluctuations increase further eventually reaching a plateau ($\frac{\Delta J^2}{L} \to t_h^2 \;, \, U \to \infty$) while the entanglement entropy decays as $\frac{1}{U^2}$ in the large interaction limit.
We note that this does not contradict our previous statements around Eq.~(\ref{eq:Implication1}): current fluctuations are a necessary condition for light-matter entanglement but not a sufficient one.

We observe qualitatively similar behavior for the photon number $N_{\rm phot}$, where the high-frequency result Eq.~(\ref{eq:nPhotHighFreq}) compares well to the DMRG result for smaller interactions. One notable difference is that at $U=0$ there is a finite photon number $N_{\rm phot}>0$ due to a contribution from the squeezing of the photon states.

The behavior at larger interaction strength, $U > 3t_{\rm h}$, of the entanglement entropy and the photon number can be understood by noting that deep in the CDW phase, applying the light-matter coupling to the ground state wave function of the $g=0$ system does not only create a photon excitation of energy $\Omega$ but also a charge excitation (through a hopping process) of energy $U$.
To connect to the large $U$ limit, we perform second-order perturbation theory in $\frac{t_{\rm}}{U}$ (see Supplemental Material \cite{Supplementary} for details) and find that in order to account for this energy we effectively need to replace $\frac{1}{\Omega^2} \rightarrow \frac{1}{(\Omega + U)^2}$ in Eq.~(\ref{eq:nPhotHighFreq}) and Eq.~(\ref{eq:S_j}) such that we obtain
\begin{equation}
\begin{aligned}
    S(\Delta J^2, U) =& \frac{-1}{1 + \chi'} \ln \left(\frac{1}{1 +\chi'} \right) -\frac{\chi'}{1 + \chi'} \ln \left(\frac{\chi'}{1 + \chi'} \right)\\[8pt]
    \chi' =& \frac{g^2}{(\Omega + U)^2} \frac{\Delta J^2 }{L}.
\end{aligned}
    \label{eq:S_jU}
\end{equation}
for the entanglement entropy and 
\begin{equation}
N_{\rm phot}(\Delta J^2, \langle T \rangle^2, U) = \frac{g^2}{(\Omega + U)^2} \frac{\Delta J^2}{L} + \frac{g^4}{(\Omega + U)^2} \frac{\langle T \rangle^2}{L^2}.
    \label{eq:nPhotHighFreqU}
\end{equation}
for the photon number.
We compare these expressions to the result obtained with DMRG for values of the interaction in the CDW phase $U > 2t_{\rm h}$ in Fig.~\ref{fig:1}c and find excellent agreement.

\paragraph*{Comparison with mean-field approach.--} 
Above we have found the necessity of current fluctuations in the fermionic system for light-matter entanglement in the coupled groundstate. In order to further elucidate the role of entanglement between light and matter for accurate results when $g \neq 0$ and $U \neq 0$, we compare the numerically exact DMRG results to a mean-field (MF) decoupling of light and matter that assumes a factorized groundstate wave function
\begin{equation}
    |\Psi \rangle = |\psi \rangle_{\rm f} \otimes | \phi \rangle_{\rm b}.
\end{equation}
Here $|\psi\rangle_{\rm f}$ is the fermionic part of the wave function while $|\phi \rangle_{\rm b}$ is the photon part.
Importantly, both the cavity and matter systems are treated exactly by exact diagonalization and DMRG respectively, and only their interplay is approximated (for details see Supplemental Material \cite{Supplementary}).

We compare the photon number as obtained with DMRG to that obtained with the MF approach in Fig.~\ref{fig:2}a.
The DMRG result displays a peak in photon number after the BKT transition as a result of the interplay of current fluctuations and localization of the fermions at larger interaction in the CDW phase.
The MF result agrees with DMRG at vanishing interaction $U=0$ but from there on decreases monotonically as a function of increasing $U$, in stark contrast to the numerically exact DMRG result.

To obtain further insight into the specific wave functions obtained with the two methods, we plot the probability $P(n_{\rm phot})$  to find $n_{\rm phot}$ photons in the groundstate at $U = 3t_{\rm h}$.
In MF one only finds even numbers of photons, $P(2n_{\rm phot}) > 0$ while $P(2n_{\rm phot} + 1) = 0$ for $n_{\rm phot} \in \mathbb{N}$.
By contrast, the DMRG result shows a finite probability to find both even and odd photon numbers.
We conclude that the MF decoupling captures the squeezing of the cavity mode (and thus $N_{\rm phot}|_{U = 0}$ matches the DMRG result) but fails to obtain the superposition of coherent states shown in Eq.~(\ref{eq:approxWavefunction}).
This is reasonable since the latter is mediated by the fluctuations of the current operator missed by the MF approach while fluctuations of the squeezing are suppressed in the TD limit also for the exact wave function such that the squeezing can be reproduced accurately.

We finally study the longest-ranged density-density correlations accessible in the $L$-site chain,
\begin{equation}\label{eq:def_C}
    C := \langle n_0 n_{\frac{L}{2}} \rangle.
\end{equation}
$C$ serves as an order parameter for the BKT transition as $C=0$ in the LL phase while $C>0$ in the CDW phase, when $L \rightarrow \infty$.
We show the result from both DMRG and the MF approach compared to that computed via DMRG in the uncoupled chain $g=0$ in Fig.~\ref{fig:2}b.
With DMRG we find cavity-increased correlations corresponding to an enhancement of the CDW over the LL phase.
The MF approach produces a qualitatively similar result, however, with increases in correlations remaining approximately an order of magnitude smaller than those observed with DMRG.
In fact we can fully understand the MF result from an effective rescaling of the hopping according to
\begin{equation}
    t_{\rm eff} = t \langle e^{i g \left(a^{\dag} + a\right)} \rangle.
    \label{eq:effectiveHopping}
\end{equation}
Since generally $t_{\rm eff} < t$ the interaction $U$ is effectively increased over the hopping leading to increased correlations. Yet, we show this heavily underestimates the way  the cavity enhances effective correlations, and a truly light-matter entangled coupled wavefunction is required to obtain the full picture. 

Finally we perform a finite-size scaling of the changes in $C$ and find a monotonic decay as the system size $L$ increases.
The reason for this vanishing influence of the cavity on the correlator in the thermodynamic limit lies in the fact that we have approximated the cavity degrees of freedom with a single bosonic mode, which has vanishing energy density in the thermodynamic limit.

\paragraph*{Discussion.--}
In this work we have shown that quantum fluctuations of the current operator are a necessary condition for light-matter entanglement between spinless fermions in an interacting quantum chain and a cavity mode.
This result is expected to hold in more general settings including higher dimensions, different forms of the interaction, and the inclusion of a spin degree of freedom. 
The reason is that for the analytical calculations performed here, neither dimensionality nor the particular form of the interaction played any role whatsoever.
We used the dipole approximation which is expected to hold in the far field in optical cavities since $c \gg v_{\rm F}$ .
This assumption might, however, be violated in plasmonic cavities and in the presence of strong near-field effects.
The fluctuations of the current operator can in principle be directly measured through Kubo response functions \cite{hauke_measuring_2016}, possibly providing a quantum-metrological handle to measure light-matter entanglement \cite{vedral_quantifying_2008, giovannetti_advances_2011}.
This is reminiscent of the recent finding that the quantum Fisher information, that is directly related to the quantum fluctuations of an operator \cite{giovannetti_advances_2011}, is a multipartite entanglement witness in \cite{hyllus_fisher_2012} and out of thermal equilibrium \cite{baykusheva_witnessing_2022}.
However, we point out that the relations found in this work establish quantum fluctuations as a necessary condition for entanglement and not a sufficient one, as would be required \cite{terhal_bell_2000} for quantum fluctuations to serve as a complete entanglement witness. We note that the role of quantum fluctuations for light-matter entanglement in cavity quantum materials at finite temperature, where also thermal fluctuations are present, and out of thermal equilibrium are interesting subjects that deserve further study.

Additionally we showed that the entanglement between light and matter can lead to increased effective correlations. Finally we demonstrated by finite-size scaling that a single cavity mode can only influence matter properties in mesoscopic, but not macroscopic systems. 
The inclusion of macroscopically many modes, that are naturally present in certain experimental setups, could for instance enable the creation of long-ranged charge correlations in a regime where correlations are otherwise only short-ranged. Such an effect might explain the recent experiment in which a Fabry-Perot resonator modified the metal-to-CDW-insulator transition in {1T}-{TaS}$_{2}$ \cite{jarc_cavity_2022}.
On the flip side, the measurement of modifications of both electronic and photonic properties in hybrid systems that cannot be understood from a product wave function alone might point us to finding entanglement in cavity-matter systems.

\section*{Data and Code availability}
Data included in the paper are available upon request,
the codes used to generate it are openly available at \url{https://github.com/GiacomoPassetti/Paper_XXZ_cavity.git}.

\section*{ACKNOWLEDGEMENTS} 
We would like to thank Gian Marcello Andolina, Valentin Bruch, Fabio Cavaliere, Titas Chanda, Giuliano Chiriacò, Pavel Dolgriev, Francesco Grandi, Mohammad Hafezi, Angel Rubio, Frank Schlawin and Lukas Weber for fruitful discussions.

We acknowledge support by the Deutsche Forschungsgemeinschaft (DFG, German Research Foundation) via  Germany’s Excellence Strategy -- Cluster of Excellence Matter and Light for Quantum Computing (ML4Q) EXC 2004/1 -- 390534769 and within the RTG 1995. We also acknowledge support from the Max Planck-New York City Center for Non-Equilibrium Quantum Phenomena. MAS acknowledges financial support through the Deutsche Forschungsgemeinschaft (DFG, German Research Foundation) via the Emmy Noether program (SE 2558/2).
Simulations were performed with computing resources granted by RWTH Aachen University under project rwth0926 and on the HPC system Cobra at the Max Planck Computing and Data Facility.

\bibliography{XXZ.bib}

\end{document}


\title{Supplemental Material for:
Cavity Light-Matter Hybridization Driven by Quantum Fluctuations}

\newcommand{\affiliationRWTH}{
Institut f\"ur Theorie der Statistischen Physik, RWTH Aachen University and JARA-Fundamentals of Future Information Technology, 52056 Aachen, Germany
}
\newcommand{\affiliationMPSD}{
Max Planck Institute for the Structure and Dynamics of Matter,
Center for Free-Electron Laser Science (CFEL),
Luruper Chaussee 149, 22761 Hamburg, Germany
}

\newcommand{\affiliationUNIGE}{
Dipartimento di Fisica, Università di Genova, 16146, Genova, Italy
}
\newcommand{\affiliationCNR}{
SPIN-CNR, 16146, Genova, Italy
}

\author{Giacomo Passetti}
\thanks{These authors contributed equally}
\affiliation{\affiliationRWTH}

\author{Christian J. Eckhardt}
\thanks{These authors contributed equally}
\affiliation{\affiliationRWTH}
\affiliation{\affiliationMPSD}

\author{Fabio Cavaliere}
\affiliation{\affiliationUNIGE}
\affiliation{\affiliationCNR}


\author{Michael~A. Sentef}
\affiliation{\affiliationMPSD}

\author{Dante M.~Kennes}
\affiliation{\affiliationRWTH}
\affiliation{\affiliationMPSD}

\date{\today}
\maketitle

\section{High-frequency expansion}

In this part we will show how to obtain the approximate form of the wave-function (Eq.~(2) of the main part) that is accurate in the high frequency limit and how to compute the photon number and the entanglement entropy with it.
We consider the Hamiltonian to second order in the light-matter coupling which gives the correct ground state in the thermodynamic limit $L \to \infty$ and if the light-matter coupling is small which is fulfilled since $g t_{\rm h} \ll \Omega$ in the here considered limit
\begin{equation}
\begin{aligned}
    H^{g^2} &= \Omega a^{\dag} a + T + H_{\rm int} + \frac{g}{\sqrt{L}} \left( a^{\dag} + a \right) J - \frac{g^2}{2 L} \left( a^{\dag} + a \right)^2 T\\
    H_{\rm int} &= U \sum_{i} \left( n_i - \frac{1}{2}\right)\left( n_{i + 1} - \frac{1}{2}\right).
\end{aligned}
\end{equation}
We perform two consecutive transformations of the Hamiltonian according to
\begin{equation}
\begin{aligned}
    \tilde{H}^{g^2} &= e^{S^{\rm sq}[\xi(T)]} e^{S^{\rm d}[\alpha(J)]} H^{g^2}  e^{-S^{\rm d}[\alpha(J)]} e^{-S^{\rm sq}[\xi(T)]}\\
    S^{\rm d}[\alpha(J)] &= \alpha(J) \left(a^{\dag} - a\right) \: ; \; \alpha(J) = \frac{g}{\Omega} \frac{J}{\sqrt{L}}\\
    S^{\rm sq}[\xi(T)] &= \frac{1}{2} \xi(T) \left(a^2 - \left(a^{\dag}\right)^2\right) \: ; \; \xi(T) = \frac{1}{2} \ln \left( 1 - 2 \frac{g^2}{\Omega} \frac{T}{L}\right)
\end{aligned}
\label{eq:trafo}
\end{equation}
which yields for $\tilde{H}$
\begin{equation}
    \tilde{H}^{g^2} = \tilde{\Omega} \beta^{\dag} \beta + T + H_{\rm int}
\end{equation}
where we have dropped terms multiplied with $\frac{1}{\Omega}$ since they will become irrelevant in the high frequency limit.
Here $\beta$ annihilates a squeezed coherent state of the cavity mode.
$\tilde{\Omega}$ is a renormalized frequency that is identical to the original frequency $\Omega$ to leading order in $\frac{1}{\Omega}$
\begin{equation}
    \tilde{\Omega} = \Omega \sqrt{1 - 2 \frac{g^2}{\Omega} \frac{\langle T \rangle}{L}} = \Omega + \mathcal{O}(\frac{1}{\Omega})
\end{equation}
such that we will keep performing computations with the original frequency $\Omega$.

In the new squeezed-coherent basis light and matter degrees of freedom decouple to leading order in $\frac{1}{\Omega}$ such that we can write the ground state wave-function as
\begin{equation}
    | \tilde{\Psi} \rangle = |\psi\rangle \otimes |0_{\beta}\rangle
\end{equation}
where $|\psi \rangle$ is the fermionic part of the wave-function and $|0_{\beta} \rangle$ is the squeezed-coherent vacuum.
In order to calculate the photon number and light-matter entanglement we need to obtain the ground state wave-function in the original basis which we do by applying the inverse transformation to Eq.~(\ref{eq:trafo}).
In order to do this we write the fermionic part of the wave-function in the common eigenbasis of $T$ and $J$ ($[T, J] = 0$) ie. in the Bloch basis
\begin{equation}
    |\tilde{\Psi} \rangle = \sum_u c_u |j_u, t_u \rangle \otimes |0_{\beta} \rangle
\end{equation}
where $c_u \in \mathbb{C}$ are complex numbers fulfilling the normalization condition $\sum_u |c_u|^2 = 1$. $t_u$ and $j_u$ denote the eigenvalue of $T$ and $J$ respectively to the corresponding Bloch state $T|j_u, t_u \rangle = t_u |j_u, t_u \rangle$ and $J | j_u, t_u \rangle = j_u |j_u, t_u \rangle$.
Here the index $u$ also counts potential degeneracies.
The ground state in the original cavity-basis can thus be computed as
\begin{equation}
    \begin{aligned}
        |\Psi \rangle &= e^{-S^{\rm d}[\alpha(J)]} e^{-S^{\rm sq}[\xi(T)]} |\tilde{\Psi} \rangle\\
        &= e^{-S^{\rm d}[\alpha(J)]} e^{-S^{\rm sq}[\xi(T)]} \sum_u c_u |j_u, t_u\rangle \otimes |0_{\beta}\rangle\\
        &= \sum_u c_u |j_u, t_u\rangle \otimes e^{-S^{\rm d}[\alpha(j_u)]} e^{-S^{\rm sq}[\xi(t_u)]} |0_{\beta}\rangle\\
        &= \sum_u c_u |j_u, t_u\rangle \otimes |\alpha(j_u), \xi(t_u) \rangle.
    \end{aligned}
    \label{eq:wavefunctionHighFreqLimit}
\end{equation}
Here $|\alpha(j_u), \xi(t_u) \rangle$ is the squeezed coherent state with coherent displacement $\alpha(j_u) = - \frac{g}{\Omega} \frac{j_u}{\sqrt{L}}$ and squeezing $\xi(t_u) = \frac{1}{2} \ln \left( 1 - 2 \frac{g^2}{\Omega} \frac{t_u}{L} \right)$.
Eq.~(\ref{eq:wavefunctionHighFreqLimit}) is the approximate form of the ground state wave-function that is accurate in the high frequency limit reported in the main part of the paper in Eq.~(2).

We continue by computing the photon number $N_{\rm phot}$ with this form of the wave-function
\begin{equation}
    \begin{aligned}
        N_{\rm phot} = \langle a^{\dag} a \rangle &= \sum_u |c_u|^2 \langle \alpha(j_u), \xi(t_u)| a^{\dag} a | \alpha(j_u), \xi(t_u) \rangle\\
        &= \sum_u |c_u|^2 \langle \xi(t_u) | a^{\dag} a + \frac{g^2}{\Omega^2} \frac{j_u^2}{L} | \xi(t_u)\rangle\\
        &= \sum_u |c_u|^2 \frac{g^2}{\Omega^2} \frac{j_u^2}{L} + \sum_u |c_u|^2 \sinh^2 \left( \frac{1}{2}  \ln\left( 1 - 2 \frac{g^2}{\Omega} \frac{t_u}{L} \right) \right)\\
        &= \frac{g^2}{\Omega^2} \frac{\langle J^2 \rangle}{L} + \frac{g^4}{\Omega^2} \frac{\langle T^2 \rangle}{L^2} + \mathcal{O}(\frac{1}{\Omega^4}).
    \end{aligned}
\end{equation}
The current operator has vanishing expectation value in the ground state $\langle J \rangle = 0$ since inversion symmetry is not broken.
On the other hand
\begin{equation}
    \frac{\langle T^2 \rangle}{L^2} = \frac{\langle T \rangle^2}{L^2} + \frac{\Delta T^2}{L^2} \to \frac{\langle T \rangle^2}{L^2} \; (L \to \infty).
\end{equation}
The fluctuation contribution from the kinetic energy term is suppressed in the thermodynamic limit (the fluctuations of the kinetic energy operator $\Delta T^2$ scale like $L$) such that we get for the photon number
\begin{equation}
    N_{\rm{phot}} = \frac{g^2}{\Omega^2} \frac{\Delta J^2}{L} + \frac{g^4}{\Omega^2} \frac{\langle T \rangle^2}{L^2}.
\end{equation}

We will next turn to computing the entanglement entropy between light and matter.
For this we compute the reduced density matrix of the photons $\rho_{\rm ph}$ by performing a partial trace over the fermionic degrees of freedom $\rm{Tr}_{\rm f}$ on the density matrix $\rho$ of the combined system as obtained from the ground state wave-function Eq.~(\ref{eq:wavefunctionHighFreqLimit})
\begin{equation}
    \rho_{\rm ph} = \text{Tr}_{\rm f} \rho = \sum_u |c_u|^2 |\alpha(j_u), \xi(t_u) \rangle \langle \alpha(j_u), \xi(t_u)|
\end{equation}
which we find to be a statistical ensemble of squeezed coherent states.
This is the reduced density matrix of the photons reported in Eq.~(4) in the main part.
We aim at calculating the entropy 
\begin{equation}
    S = -\rm{k}_{\rm B} \rm{Tr} \rho_{\rm ph} \ln \rho_{\rm ph}.
    \label{eq:entropy}
\end{equation}
From now we set $\rm{k}_{\rm B} = 1$.
The intuition for the computation is that the squeezing part of the wave-function does not contribute significantly to the entropy since it is centered around the expectation value of the squeezing
\begin{equation}
    \xi_0 = \frac{1}{2}\ln\left(1 - 2 \frac{g^2}{\Omega} \frac{\langle T \rangle }{L}\right)
\end{equation}
while fluctuations around that value are suppressed in the thermodynamic limit as previously seen in the case of the photon number.
To explicitly see this, we insert a $1$ into the $\rm Tr$ Eq.~(\ref{eq:entropy}) and use the cyclic property of the $\rm Tr$
\begin{equation}
    \begin{aligned}
        S &= - \text{Tr} e^{S^{\rm sq}[\xi_0]} e^{-S^{\rm sq}[\xi_0]} \rho_{\rm ph} \ln \rho_{\rm ph}\\
        &= - \rm{Tr} \tilde{\rho}_{\rm ph} \ln \tilde{\rho}_{\rm ph}
    \end{aligned}
\end{equation}
where
\begin{equation}
    \tilde{\rho}_{\rm ph} = \sum_u |c_u|^2 |\alpha(j_u), \xi(t_u) - \xi_0\rangle \langle \alpha(j_u), \xi(t_u) - \xi_0 |.
\end{equation}
One can write this using the squeezing transformation and writing the squeezing factor $\xi$ to leading order in $\frac{1}{\Omega}$ as
\begin{equation}
    \tilde{\rho}_{\rm ph} = \sum_u |c_u|^2 e^{\frac{1}{8} \left( a^2 - \left(a^{\dag}\right)^2 \right) \frac{t_u - \langle T \rangle}{L}}  |\alpha(j_u) \rangle \langle \alpha(j_u) | e^{-\frac{1}{8} \left( a^2 - \left(a^{\dag}\right)^2 \right) \frac{t_u - \langle T \rangle}{L}}   .
\end{equation}
From expanding the exponentials it becomes clear that the leading order contribution to the matrix elements of $\tilde{\rho}_{\rm ph}$ from the squeezing will be proportional to $\frac{\langle \left(T - \langle T \rangle \right)^2 \rangle}{L^2} = \frac{\Delta T^2}{L^2} \to 0 \; (L \to \infty)$ and therefore vanish in the thermodynamic limit.
We can therefore discard the squeezing for the computation of the entanglement entropy and instead write
\begin{equation}
    S = - \text{Tr} \rho'_{\rm ph} \ln \rho'_{\rm ph} \:;\; \rho'_{\rm ph} = \sum_u |c_u|^2 |\alpha(j_u)\rangle \langle \alpha(j_u) |.
\end{equation}
This is the form of the density matrix used in the main part Eq.~(5) to argue about current fluctuations as a necessary condition for light-matter entanglement.
Writing the coherent states in their photon number representation one obtains
\begin{equation}
\begin{aligned}
    \rho'_{\rm ph} &= \sum_u |c_u|^2 e^{-|\alpha(j_u)|^2} \sum_{n, m} \frac{\alpha(j_u)^n}{\sqrt{n!}} \frac{(\alpha(j_u)^{*})^m}{\sqrt{m!}} |n\rangle \langle m |\\
    &\overset{j_u\in\mathbb{R}}{=} \sum_u |c_u|^2 e^{-\frac{g^2}{\Omega^2}\frac{j_u^2}{L}} \sum_{n, m} \frac{(-\frac{g}{\Omega}\frac{j_u}{\sqrt{L}})^n}{\sqrt{n!}} \frac{(-\frac{g}{\Omega} \frac{j_u}{\sqrt{N}})^m}{\sqrt{m!}} |n\rangle \langle m |.
\end{aligned}
\end{equation}
Thus the matrix elements of the reduced density matrix of the photons read, in the photon number basis
\begin{equation}
    \langle n |\rho'_{\rm ph} | m \rangle = \sum_u |c_u|^2 e^{-\frac{g^2}{\Omega^2}\frac{j_u^2}{L}} \sum_{n, m} \frac{(-\frac{g}{\Omega}\frac{j_u}{\sqrt{L}})^{n + m}}{\sqrt{n!}\sqrt{m!}}.
\end{equation}
We can relate this quantity to moments of the current operator.
For this we first note
\begin{equation}
\begin{aligned}
    &\langle \Psi | e^{-\frac{g^2}{\Omega^2}\frac{J^2}{L}} \frac{\left(-\frac{g}{\Omega}\frac{J}{\sqrt{L}}\right)^{(n + m)}}{\sqrt{n!}\sqrt{m!}}| \Psi \rangle\\
    =& \sum_{u, v} c_u^{*} c_v \langle j_u| \otimes \langle \alpha_u|  e^{-\frac{g^2}{\Omega^2}\frac{J^2}{L}} \frac{\left(-\frac{g}{\Omega}\frac{J}{\sqrt{L}}\right)^{(n + m)}}{\sqrt{n!}\sqrt{m!}} |j_v \rangle \otimes | \alpha_v \rangle\\
    =& \sum_{u} |c_u|^2 e^{-\frac{g^2}{\Omega^2}\frac{j_u^2}{L}} \frac{\left(-\frac{g}{\Omega}\frac{j_u}{\sqrt{L}}\right)^{(n + m)}}{\sqrt{n!}\sqrt{m!}}.
\end{aligned}
\end{equation}
This shows that in the high frequency limit, the entanglement entropy between light and matter is completely determined by the moments of the current operator together with the light-matter coupling $g$ and the frequency $\Omega$.
When assuming separability of higher order correlation functions as $\langle J^{2n} \rangle \approx \langle J^2 \rangle^n$ (keeping in mind that $\langle J^{2n + 1} \rangle = 0, \, n \in \mathbb N$ due to symmetry) we can also write
\begin{equation}
    \langle e^{-\frac{g^2}{\Omega^2}\frac{J^2}{L}} \frac{\left(-\frac{g}{\Omega}\frac{J}{\sqrt{L}}\right)^{(n + m)}}{\sqrt{n!}\sqrt{m!}} \rangle \approx e^{-\frac{g^2}{\Omega^2}\frac{\langle J^2 \rangle}{L}} \frac{\left(-\frac{g}{\Omega} \right)^{(n + m)}\frac{\langle J^{(n + m)} \rangle}{\sqrt{L}^{(n + m)}}}{\sqrt{n!}\sqrt{m!}}.
\end{equation}
Therefore we write the matrix elements of the reduced density matrix as
\begin{equation}
     \langle n |\rho'_{\rm ph} | m \rangle \approx e^{-\frac{g^2}{\Omega^2}\frac{\langle J^2 \rangle}{L}} \frac{\left(-\frac{g}{\Omega} \right)^{(n + m)}\frac{\langle J^{(n + m)} \rangle}{\sqrt{L}^{(n + m)}}}{\sqrt{n!}\sqrt{m!}}.
\end{equation}
Since we are assuming $\Omega \gg t_{\rm h}, U$, we can limit ourselves to the lowest order photon number states for calculating the entanglement entropy
\begin{equation}
    \begin{aligned}
    &\rho'_{\rm ph} = e^{-\frac{g^2}{\Omega^2}\frac{\langle J^2 \rangle}{L}} \left( |0 \rangle \langle 0 | + \frac{g^2}{\Omega^2} \frac{\langle J^2 \rangle}{L} |1\rangle \langle 1| \right) + \mathcal{O}\left( \frac{g^4 t_{\rm h}^4}{\Omega^4} \right)\\
    &= \frac{1}{1 + \frac{g^2}{\Omega^2} \frac{\langle J^2 \rangle}{L}} \left( |0 \rangle \langle 0 | + \frac{g^2}{\Omega^2} \frac{\langle J^2 \rangle}{L} |1\rangle \langle 1| \right) + \mathcal{O}\left( \frac{g^4 t_{\rm h}^4}{\Omega^4} \right).
    \end{aligned}
    \label{eq:photonDensityMatrixFinal}
\end{equation}
In principle, one would also need to consider elements of the density matrix of the form $|0\rangle \langle 2|$ but one can show that they don't contribute to the entropy in our level of approximation.
The $\mathcal{O}$ notation is to be understood in the operator norm here.
The entropy of the density matrix Eq.~(\ref{eq:photonDensityMatrixFinal}) is
\begin{equation}
\begin{aligned}
    S_{\rm ph} =& -\frac{1}{1 + \frac{g^2}{\Omega^2} \frac{\langle J^2 \rangle}{L}} \ln \left(\frac{1}{1 + \frac{g^2}{\Omega^2} \frac{\langle J^2 \rangle}{L}} \right) \\
    &-\frac{\frac{g^2}{\Omega^2} \frac{\langle J^2 \rangle}{L}}{1 + \frac{g^2}{\Omega^2} \frac{\langle J^2 \rangle}{L}} \ln \left(\frac{\frac{g^2}{\Omega^2} \frac{\langle J^2 \rangle}{L}}{1 + \frac{g^2}{\Omega^2} \frac{\langle J^2 \rangle}{L}} \right)
\end{aligned}
\end{equation}
which matches Eq.~(7) of the main part.

\section{Numerical convergence}\label{sec:numerics}
\begin{figure*}[t]
    \centering
    \includegraphics[scale= 1]{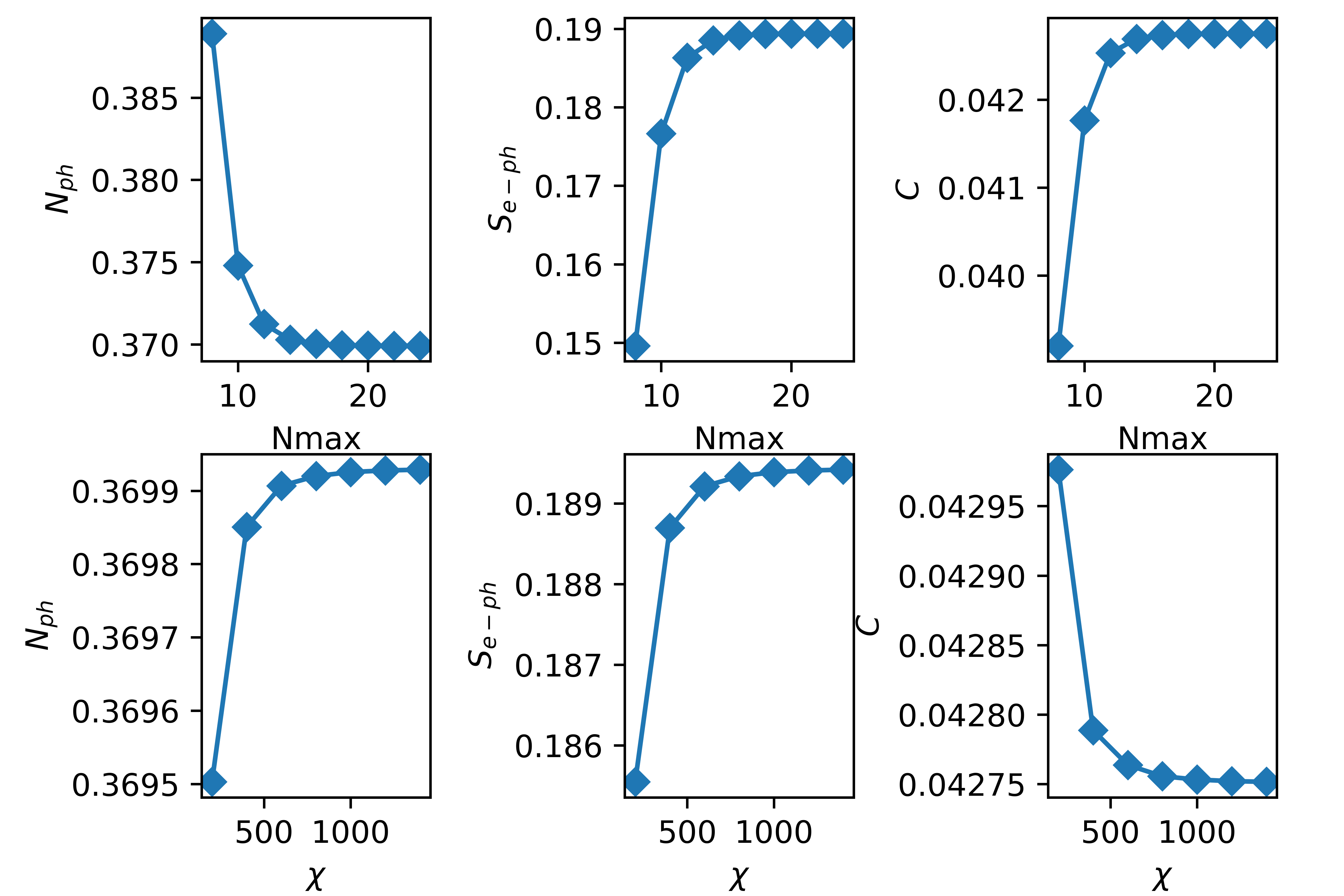}%
    \caption{Convergence analysis for the relevant quantities $N_{\rm{ph}}$, $S_{\rm{e-ph}}$ and $C$. All the curves have been obtained setting $L = 110$, $U = 2.4t_{\rm h}$, $g = 4$ and $\Omega = 1t_{\rm h}$.}
    \label{fig:convergence}
\end{figure*}
Here we report convergence checks for our DMRG results.
There are two parameters that set the limit of the accuracy of our numerical approximation, which are photon cutoff $N_{max}$ and the bond dimension $\chi$.
The photon cutoff sets the dimension of the local Hilbert space that describes the bosonic degree of freedom, which we considered to be finite.
The maximum bond dimension $\chi$ defines the maximum dimension for the virtual bonds in the MPS representation of the ground state wavefunction \cite{schollwock_density-matrix_2005}.
The plots reported in Fig.~\ref{fig:convergence} have all been obtained simulating a system of size $L = 110$ for $U = 2.4t_{\rm h}$ and $g = 4$, which relatively to the typical parameters discussed in the paper corresponds to high value of the coupling in the critical region. 
We show the convergence behaviour for three significant quantities discussed in the main text, nominally $N_{\rm{ph}}$, $S_{\rm{e{-}ph}}$ and $C$. 
Unless otherwise stated, the DMRG parameters for the results shown in the paper are $\chi = 1000$ and $N_{max} = 16$.

\section{Connection to high $U$ limit via Perturbation theory}
\label{sec:perturbationTheory}
In the main part we noted, that in the CDW phase one needs to account for the fact that the light-matter coupling does not only create a light-excitation by creating a photon but also a charge excitation through a fermionic hopping.
In this part we will show that this essentially corresponds to a replacement of $\frac{1}{\Omega^2} \to \frac{1}{(\Omega + U)^2}$ in the formulas found in the high frequency limit.
To this end, we will consider the large $U$ limit and perform second order perturbation theory in $\frac{t_{\rm h}}{U}$ again only keeping terms including the second order light-matter coupling.
We thus define
\begin{equation}
    \begin{aligned}
        H_0 &= \Omega a^{\dag} a + U \sum_j \left(n_j - \frac{1}{2}\right) \left( n_{j + 1} - \frac{1}{2} \right)\\
        V &=T + \frac{g}{\sqrt{L}} \left( a^{\dag} + a \right) J - \frac{g^2}{2 L} \left(a^{\dag} + a\right)^2 T.
    \end{aligned}
    \label{eq:splitUpH}
\end{equation}
where the definitions of $T$ (kinetic energy operator) and $J$ (current operator) are in the main part.
Denoting by $|\psi_n^{(0)} \rangle$ the eigenstates of the unperturbed system $H_0$ with eigenenergy $E_n$ one can write the GS of the perturbed system to second order in $t_{\rm h}$ as 

\begin{equation}
    \begin{aligned}
        |\psi_0^{(2)} \rangle &= |\psi_0^{(0)} \rangle + \sum_{n > 0} \frac{\langle \psi_n^{(0)} | V | \psi_0^{(0)} \rangle}{E_0 - E_n} | \psi_n^{(0)} \rangle + \\
        &+ \sum_{n, m > 0} \frac{\langle \psi_n^{(0)} | V | \psi_m^{(0)} \rangle \langle \psi_m^{(0)} | V | \psi_0^{(0)} \rangle}{(E_n - E_m)(E_0 - E_m)} | \psi_n^{(0)} \rangle \\
        &- \sum_{n > 0} \frac{\langle \psi_n^{(0)} | V | \psi_0^{(0)} \rangle \langle \psi_0^{(0)} | V | \psi_0^{(0)} \rangle}{(E_n - E_0)^2} | \psi_n^{(0)} \rangle + \\
        &-\frac{1}{2} \sum_{n > 0} \frac{|\langle \psi_n^{(0)} | V | \psi_0^{(0)} \rangle|^2}{(E_0 - E_n)^2} | \psi_0^{(0)} \rangle.
        \label{eq:stateSecondOrder}
    \end{aligned}
\end{equation}
With this state one can calculate the photon number to second order in $t_{\rm h}$ as
\begin{equation}
\begin{aligned}
    N_{\rm ph}^{(2)} &= \langle \psi_0^{(2)} | a^{\dag} a | \psi_0^{(2)} \rangle = \\
    &= \sum_{n, m > 0} \frac{\langle \psi_0^{(0)} | V^{\dag} | \psi_m^{(0)} \rangle \langle \psi_n^{(0)} | V | \psi_0^{(0)} \rangle }{(E_0 - E_n) (E_0 - E_m)} \langle \psi_m^{(0)} | a^{\dag} a | \psi_n^{(0)} \rangle\\
    &= \sum_{n > 0} \frac{\langle \psi_0^{(0)} | V^{\dag} | \psi_n^{(0)} \rangle \langle \psi_n^{(0)} | V | \psi_0^{(0)} \rangle }{(E_0 - E_n)^2} \langle \psi_n^{(0)} | a^{\dag} a | \psi_n^{(0)} \rangle
    \label{eq:photNumPerturb}
\end{aligned}
\end{equation}
where in the second line we assumed the basis states to be written in the photon number basis and used the fact that the operator $a^{\dag} a$ is diagonal in that basis.
Noting that only the second and third term in $V$ (Eq.~(\ref{eq:splitUpH})) create a non-zero photon number and $\langle \psi_0^{(0)} | V | \psi_0^{(0)} \rangle = 0$ we can write Eq.~(\ref{eq:photNumPerturb}) as
\begin{equation}
    \begin{aligned}
        N_{\rm ph}^{(2)} = \frac{1}{(U + \Omega)^2}  \langle \psi_0^{(0)} | \left( \frac{g}{\sqrt{L}} \left( a^{\dag} + a \right) J - \frac{g^2}{2L} \left( a^{\dag} + a \right)^2 T\right)^2 | \psi_0^{(0)} \rangle.
    \end{aligned}
    \label{eq:nPhotPerturb1}
\end{equation}
The GS of the unperturbed system $|\psi_0^{(0)} \rangle$ can be written as
\begin{equation}
    |\psi_0^{(0)} \rangle = | \psi^{\rm{CDW}} \rangle \otimes |0 \rangle
\end{equation}
where $|0\rangle$ is the photon vacuum state and $| \psi^{\rm{CDW}} \rangle$ is the state with staggered occupation (or the symmetric superposition of the two possible states in the case of no explicit symmetry breaking).
Inserting this into Eq.~(\ref{eq:nPhotPerturb1}) we obtain
\begin{equation}
\begin{aligned}
     N_{\rm ph}^{(2)} = &\frac{g^2}{L(U + \Omega)^2}  \langle 0 | \left( a^{\dag} + a \right)^2| 0 \rangle \langle \psi^{\rm{CDW}} | J^2 | \psi^{\rm{CDW}} \rangle + \\&
     \frac{g^4}{4L^2(U + \Omega)^2}  \langle 0 | \left( a^{\dag} + a \right)^4| 0 \rangle \langle \psi^{\rm{CDW}} | T^2 | \psi^{\rm{CDW}} \rangle 
     - \\&
     \frac{g^3}{\sqrt{L}^3 (U + \Omega)^2}  \langle 0 | \left( a^{\dag} + a \right)^3| 0 \rangle \langle \psi^{\rm{CDW}} | J \, T | \psi^{\rm{CDW}} \rangle.
\end{aligned}
\label{eq:nPhotPerturb}
\end{equation}
The last term vanishes due to the uneven number of photon creation and annihilation operators.
The other two require the calculation of the $\langle J^2 \rangle$ and $\langle T^2 \rangle$ in the CDW state.
We find
\begin{equation}
\begin{aligned}
    &\langle \psi^{\rm CDW} | J^2 | \psi^{\rm CDW} \rangle = \langle \psi^{\rm CDW} | T^2 | \psi^{\rm CDW} \rangle = \\
    &- t_{\rm h}^2 \sum_{i, j} \delta_{i, j} \langle \psi | c_{i + 1}^{\dag} c_{i}  c_{j}^{\dag} c_{j + 1} +  c_{i}^{\dag} c_{i + 1}  c_{j + 1}^{\dag} c_{j} | \psi \rangle = L t_{\rm h}^2 .
\end{aligned}
\label{eq:CurrentFluctuationsLargeU}
\end{equation}
Thus the second term in Eq.~(\ref{eq:nPhotPerturb}) also vanishes in the thermodynamic limit $L \to \infty$.
This is consistent with our observations in the main part since in the CDW state $\langle T \rangle = 0$ and the contribution from fluctuations is suppressed in the large $L$ limit as seen explicitly here.
Thus we obtain the same relation of the photon number to the current fluctuations as in the high frequency limit (Eq.~(3) of the main part) but with the replacement $\frac{1}{\Omega^2} \rightarrow \frac{1}{(\Omega + U)^2}$.

Overall we get for the photon number in the large $U$ limit
\begin{equation}
    N_{\rm ph}^{(2)} = \frac{g^2 t_h^2}{(U + \Omega)^2}.
\end{equation}
This matches precisely the results obtained with DMRG for large interaction values $U \gg t_{\rm h}$ and $U \gg \Omega$.

\subsection{Entanglement entropy accounting for charge excitations in the CDW}
We calculate the entanglement entropy to second (leading) order in $\frac{t_{\rm h}}{U}$ in perturbation theory. 
To this end we split up the perturbative wave-function as
\begin{equation}
\begin{aligned}
    | \phi \rangle =& \left(c_0 |\phi_0 \rangle + c_1^0 |\phi_1^0 \rangle + c_2^0 |\phi_2^0 \rangle \right) \otimes | 0 \rangle + \left(c_1^1 |\phi_1^1 \rangle +
    + c_2^1 |\phi_2^1 \rangle \right) \otimes | 1 \rangle 
    + \left(c_1^2 |\phi_1^2\rangle  c_2^2 |\phi_2^2 \rangle \right) \otimes | 2 \rangle\\ 
    &+ c_2^3 |\phi_2^3 \rangle \otimes | 3 \rangle + + c_2^4 |\phi_2^4 \rangle \otimes | 4 \rangle.
    \end{aligned}
\end{equation}
Here $|\phi_0 \rangle$ denotes the fermionic part of the unperturbed GS which in the large $U$ limit is the staggered-occupied state. 
For the coefficients $c_m^n$ and the fermionic wave-functions $|\phi_m^n \rangle$, $m$ always denotes the order in perturbation theory that one gets by repeatedly applying $J$ and $T$ to the GS and $n$ denotes the photon number that only increases when applying terms that contain photon creators or annihilators.
In order to obtain the reduced density matrix for the photons we need to calculate several scalar products of the fermionic states.
Writing only those that don't vanish or are of higher order in the perturbation we get for the reduced density matrix
\begin{equation}
\begin{aligned}
    \rho_{\rm ph} &= \rm{Tr}_{e^-} \rho = \underbrace{\left( |c_0|^2 + c_0 \left(c_2^{0}\right)^* \langle \phi_2^0 | \phi_0 \rangle + c_0^* c_2^0 \langle \phi_0 | \phi_2^0 \rangle \right)}_{=:\kappa_0} |0 \rangle \langle 0| +\underbrace{|c_1^1|^2}_{=:\kappa_1} |1 \rangle \langle 1| + \underbrace{|c_1^2|^2}_{:=\kappa_2} | 2 \rangle \langle 2 | \\&
    + \underbrace{\left(c_0 \left(c_2^2 \right)^{*}  \langle \phi_2^2 | \phi_0 \rangle + c_1^0 \left(c_1^2\right)^* \langle \phi_1^2 | \phi_1^0 \rangle \right)}_{=:\kappa_{02}} |0\rangle \langle 2| + \underbrace{ \left( c_0^* c_2^{2}  \langle \phi_0 | \phi_2^0 \rangle + \left(c_1^0\right)^* c_1^2 \langle \phi_1^0 | \phi_1^2 \rangle \right)}_{=:\kappa_{02}^*} |2 \rangle \langle 0 |.
\end{aligned}
\label{eq:rhoPerturb1}
\end{equation}
Written in matrix form this reads
\begin{equation}
    \rho_{\rm ph} = \begin{pmatrix}
    \kappa_0 & 0 & \kappa_{02}\\
    0 & \kappa_1 & 0\\
    \kappa_{02}^* & 0 & \kappa_2
    \end{pmatrix}
    \overset{\rm reorder}{\rightarrow}
    \begin{pmatrix}
    \kappa_0 & \kappa_{02} & 0\\
    \kappa_{02}^* & \kappa_2 & 0\\
     0 & 0 & \kappa_1
    \end{pmatrix}
\end{equation}
As seen in the previous section, $\kappa_2$ involves the expectation value $\frac{1}{L^2} \langle \psi^{\rm{CDW}} | T^2 | \psi^{\rm{CDW}} \rangle$ and therefore $\kappa_2 \to 0$ for $L \to \infty$.
To find the entropy we diagonalize the remaining matrix which we can now do by blocks -- in fact we only need to diagonalize the upper $2\times 2$ block.
The eigenvalues read
\begin{equation}
\begin{aligned}
    v_{\pm} =& \frac{1}{2} \left( \kappa_0 \pm \sqrt{\kappa_0^2 + 4 |\kappa_{02}|^2} \right) = \frac{1}{2} \left( \kappa_0 \pm \kappa_0 \sqrt{1 + 4\frac{ |\kappa_{02}|^2}{\kappa_0^2}} \right).
\end{aligned}
\end{equation}
Since $\kappa_0 \sim 1$ and $\kappa_{02} \sim \left(\frac{t_{\rm h}}{U}\right)^2$ (as can be seen from Eq.~(\ref{eq:rhoPerturb1})) we can expand the square root and conclude that only one eigenvalue $\kappa_0$ is different from zero within our level of approximation.

We now remember that we have calculated $\kappa_1 = \frac{g^2 t_{\rm h}^2}{(U + \Omega)^2}$ before and fix $\kappa_0$ by normalization such that we get for the entropy
\begin{equation}
\begin{aligned}
    -S =& \left(1 - \frac{g^2 t_{\rm h}^2}{(U + \Omega)^2}\right) \ln \left(1 - \frac{g^2 t_{\rm h}^2}{(U + \Omega)^2}\right) + \frac{g^2 t_{\rm h}^2}{(U + \Omega)^2} \ln \left( \frac{g^2 t_{\rm h}^2}{(U + \Omega)^2} \right) + \mathcal{O}\left(\frac{ t_{\rm h}^3}{U^3}\right).\\
\end{aligned}
\end{equation}

This matches to leading order in $\frac{1}{U}$ the result obtained in the high-frequency limit when replacing $\frac{1}{\Omega^2} \rightarrow \frac{1}{(\Omega + U)^2}$ and inserting the current fluctuations in the large $U$ according to Eq.~(\ref{eq:CurrentFluctuationsLargeU}).

\section{Cavity-fermions Mean Field decoupling }\label{sec:mean field decoupling}
We start from the full model, defined as 
\begin{eqnarray}
    H = -t_{h}\sum_{i}\left(e^{i\frac{g}{\sqrt{L}}(\aplusa)}c^{\dag}_{i}c_{i+1}+h.c.\right) + \\ \nonumber +U\sum_{i}\left(n_{i}-\frac{1}{2}\right)\left(n_{i+1}-\frac{1}{2}\right)+ \Omega \ada
    \end{eqnarray}
We further define
\begin{equation}
    K_{L}=\sum_{i}c^{\dag}_{i}c_{i+1} \quad K_{R}=\sum_{i}c^{\dag}_{i+1}c_{i}.
\end{equation}
The single photonic mode
can be decoupled from the chain with a mean-field approach, by setting to zero the fluctuations 
\begin{equation}
    \begin{cases}
    \left(K_{R} - \braket{K_{R}}\right)\left(e^{ig(\aplusa)}-\braket{e^{ig(\aplusa)}}\right) = 0 \\
    \left(K_{L} - \braket{K_{L}}\right)\left(e^{-ig(\aplusa)}-\braket{e^{-ig(\aplusa)}}\right) = 0.
    \end{cases}
\end{equation}
This leads to the decoupled hamiltonian
\begin{eqnarray}
    H =-t_{h}\left( \braket{e^{ig(\aplusa)}}K_{R} + \braket{e^{-ig(\aplusa)}}K_{L}\right) + \\
    \nonumber + 
    U \sum_{i}\left(n_{i}-\frac{1}{2})(n_{i+1}-\frac{1}{2}\right) + 
    \\ +\Omega \ada -t_{h}\left( e^{ig(\aplusa)}\braket{K_{R}} + e^{-ig(\aplusa)}\braket{K_{L}}\right) \nonumber \\
    + t_{h}\left(\braket{K_{R}}\braket{e^{ig(\aplusa)}} + \braket{K_{L}}\braket{e^{-ig(\aplusa)}}\right)  \nonumber
\end{eqnarray}


Now we can observe that one can always write
\begin{equation}
      \braket{e^{\pm ig(\aplusa)}} = \gamma e^{\pm i\Gamma} 
\end{equation}
And that in general:
\begin{equation}
\braket{K_{R}} = \kappa e^{i\chi} = \braket{K_{L}}^*.
\end{equation}

Considering only the XXZ model, for the couple of parameters $(t, U)$ we can get the ground state that will be labeled as $\braket{GS_{t,U}}$ satisfying $H^{XXZ}_{t,U}\ket{GS_{t,U}} = E^{GS}_{t,u}$.

To shorten the notation we define 
 $t_{eff}=\gamma t_{h} e^{i\Gamma}$
and
\begin{eqnarray}\label{H_eff}
    H_{f}(\gamma, \phi) = -t_{eff}\sum_{i}\left(c^{\dagger}_{i}c_{i+1} + h.c.\right) +  \\
    \nonumber U \sum_{i}\left(n_{i}-\frac{1}{2}\right)\left(n_{i+1}-\frac{1}{2}\right) \\
    \nonumber    H_{b} =  \Omega \ada -2t_{h}\kappa\cos\left(g(a+a^{\dag})+\chi\right).
\end{eqnarray}

that allows to write the MF hamiltonian as 
\begin{equation}
    H_{MF} = H_{f} + H_{b}+ 2 t_{h} \kappa\gamma\cos\left(\chi + \phi \right).
\label{eq:hmf}
\end{equation}
The minimization of the energy will be then obtained through a numerical procedure that, starting from the solution for the uncoupled model, iteratively diagonalize the bosonic and the fermionic part until one gets a wavefunction:
\begin{equation}
     \ket{\Psi} = \ket{\psi}_{f}\otimes \ket{\phi}_{b},
\end{equation}
tensor product of a fermionic wavefunction $\ket{\psi}_{f}$ and a bosonic wavefunction $\ket{\phi}_{b}$, that satisfy the self-consistency conditions:
\begin{eqnarray}
    \braket{\Psi|K_{R}|\Psi} = \kappa \text{e}^{i\chi}  \\
    \braket{\Psi|\text{e}^{ig(a+a^{\dag})}|\Psi} = \gamma \text{e}^{ i \Gamma} \\
    H_{MF}(\gamma, \phi, \kappa, \chi) \ket{\Psi}=E  (\gamma, \phi, \kappa, \chi) \ket{\Psi}.
\end{eqnarray}

\bibliography{XXZ.bib}